\newcommand \Mpc {h_{70}^{-1}{\rm Mpc}}
\newcommand \kpc {h_{70}^{-1}{\rm kpc}}

\newcommand \arcs{\hbox{$^{\prime\prime}$}}

\newcommand \kms {{\rm km~s}^{-1}}

\newcommand \beqn {\begin{equation}}
\newcommand \eeqn {\end{equation}}

\documentclass[apjl]{emulateapj}

\begin{document}

\title{Spectroscopic Determination of the Luminosity Function in The Galaxy Clusters A2199 and Virgo}
\shorttitle{Faint End of the Luminosity Function}
\shortauthors{Rines \& Geller}

\author{Kenneth Rines and Margaret J.~Geller}
\email{krines@cfa.harvard.edu}
\affil{Smithsonian Astrophysical Observatory, 60 Garden St, MS 20, Cambridge, MA 02138; krines@cfa.harvard.edu}

\begin{abstract}

We report a new determination of the faint end of the galaxy
luminosity function in the nearby clusters Virgo and Abell 2199 using
data from SDSS and the Hectospec multifiber spectrograph on the MMT.
The luminosity function of A2199 is consistent with a single Schechter
function to $M_r=-15.6$ + 5 log $h_{70}$ with a faint-end slope of
$\alpha=-1.13\pm0.07$ (statistical).  The LF in Virgo extends to
$M_r\approx-13.5\approx M^*+8$ and has a slope of
$\alpha=-1.28\pm0.06$ (statistical).  The red sequence of cluster
members is prominent in both clusters, and almost no cluster galaxies
are redder than this sequence.  A large fraction of photometric
red-sequence galaxies lie behind the cluster.  We compare our results
to previous estimates and find poor agreement with estimates based on
statistical background subtraction but good agreement with estimates
based on photometric membership classifications (e.g., colors,
morphology, surface brightness).  We conclude that spectroscopic data
are critical for estimating the faint end of the luminosity function
in clusters.  The faint-end slope we find is consistent with values
found for field galaxies, weakening any argument for environmental
evolution in the relative abundance of dwarf galaxies.  However, dwarf
galaxies in clusters are significantly redder than field galaxies of
similar luminosity or mass, indicating that star formation processes
in dwarfs do depend on environment.

\end{abstract}


\section{Introduction}

The luminosity function of galaxies is fundamental to understanding
galaxy formation and evolution.  The luminosity function differs
dramatically from the expected mass function of dark matter halos,
indicating that baryonic physics is very important for understanding
galaxies.  In particular, a well-determined luminosity function
enables accurate modeling linking the masses of dark matter haloes to
galaxy luminosities \citep[e.g.,][and references
therein]{vale06,yang07}.  These empirical models provide a powerful
test of any model of galaxy formation and evolution.  

Early studies of
the luminosity function used the large galaxy density in clusters as a
tool for measuring the shape of the luminosity function
\citep[e.g.,][]{sandage85}.  The obvious drawback of this method is
that the luminosity function in dense environments may differ from
that in more typical galaxy environments
\citep[][]{binggeli88,driver94,depropris95}.  Environmental trends in
the luminosity function may reflect differences in galaxy formation in
different environments
\citep[][]{tully02,benson03}. For instance, tidal stripping or
``threshing'' of larger galaxies may produce dwarf galaxies
\citep{bekki01}, or dwarf galaxies may be formed in tidal tails of
intractions among giant galaxies \citep{barnes92}.  Alternatively, the
denser environments of protoclusters may have shielded low-mass
galaxies from the ultraviolet radiation responsible for reionization
\citep{tully02,benson03}.

Many studies suggest an environmental influence
on the LF; others provide no such evidence.  The main
difficulty in resolving this important issue is the challenge of
determining cluter membership for faint galaxies where background
galaxy counts are large.  

Because few deep spectroscopic surveys of
clusters extend into the dwarf galaxy regime \citep[$M_r\gtrsim$-18;
for exceptions, see][and references
therein]{mobasher03,christlein03,mahdavi05}, cluster membership is
usually determined via statistical subtraction of background galaxies
\citep[e.g.,][and references therein]{popesso06b,jenkins07,milne07,adami07,yamanoi07,barkhouse07}.  Because
galaxy number counts increase much more steeply than cluster member
counts (even for very steep faint-end slopes), small systematic
uncertainties in background subtraction can produce large
uncertainties in the abundance of faint cluster galaxies.

Here, we use MMT/Hectospec spectroscopy and data from the Sloan
Digital Sky Survey \citep[SDSS,][]{sdss} to estimate the luminosity
function (LF) in the clusters Abell 2199 and Virgo.  These data enable
very deep sampling of the luminosity function.  In particular, we
report an estimate of the faint-end slope of the luminosity function
with much smaller systematic uncertainties than most previous
investigations.  We demonstrate that photometric properties of
galaxies such as color and surface brightness correlate well with
cluster membership (in agreement with many previous studies).  Very
few galaxies redder than the red sequence are cluster members.

We discuss the photometric and spectroscopic data in $\S 2$.  We present the luminosity functions in $\S 3$.  We compare our results to previous studies and discuss possible systematic effects and uncertainties in $\S 4$.  We conclude in $\S 5$.  An Appendix details the construction of our catalog of confirmed and probable Virgo cluster members.

We assume cosmological parameters of $\Omega_m$=0.3,
$\Omega_\Lambda$=0.7, $H_0$=70~$h_{70} \kms \mbox{Mpc}^{-1}$.  The spatial
scale at the distance of A2199 is 1\arcs=0.61~$\kpc$ and
1\arcs=0.080~$\kpc$ at the distance of Virgo.


\section{Observations}

\subsection{Abell 2199}

The nearby X-ray cluster Abell 2199 \citep[e.g.,][and references
therein]{rines02} offers an excellent opportunity for probing the LF
in a rich, nearby cluster.  A2199 is significantly more massive than
Virgo \citep{cirsi} and X-ray data suggest that it is a relaxed
cluster \citep{mvfs}. The center of A2199 is dominated by NGC 6166, a
massive cD galaxy \citep{kelson02}.  

\subsubsection{Optical Imaging}

Cluster galaxies display a well-defined red sequence in
color-magnitude diagrams \citep{visvanathan77}.  Cluster
mmebers are unlikely to have colors redder than the red sequence
unless they are very dusty or have very unusual stellar populations.

\begin{figure}
\figurenum{1}
\label{rs}
\plotone{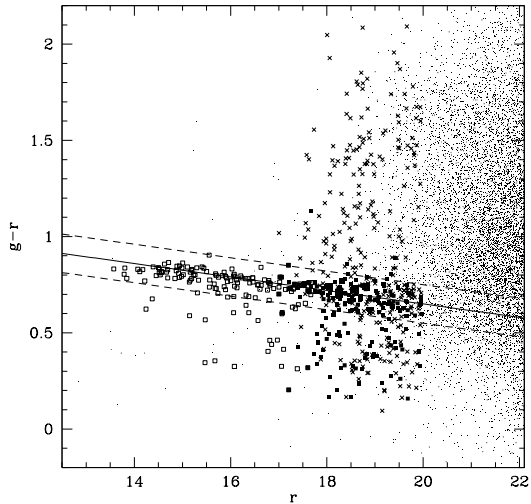}
\caption{Color-magnitude diagram for A2199.  The red sequence 
is clearly visible (solid line).  Small dots denote galaxies with SDSS
photometry and no spectroscopy.  Squares indicate spectroscopically
confirmed A2199 members from Hectospec (filled) and SDSS (open), and
crosses indicate spectroscopically confirmed background
galaxies.  Dashed lines indicate the color-magnitude cuts we adopt for
the red sequence of A2199.  Note that essentially all galaxies redder
than the red-sequence with spectra are background galaxies. }
\end{figure}

Using photometric data from SDSS, the red sequence of cluster galaxies
is readily apparent in A2199 (Figure \ref{rs}).  The red sequence can
be characterized as $g-r = -0.035(r-12)+1.0$ (solid line in Figure
\ref{rs}). Among bright galaxies ($r$$\lesssim$16) with measured
redshifts \citep[SDSS,][]{rines02}, most photometric red-sequence
galaxies are cluster members.  As recommended in the SDSS web pages,
we use composite model magnitudes as the best estimates of the galaxy
magnitudes.  Composite model magnitudes are a linear combination of
the best-fit deVaucouleurs and exponential profiles.  We correct all
magnitudes for Galactic extinction.

\subsubsection{Optical Spectroscopy}

For MMT spectroscopy, we use SDSS photometry to identify
candidate cluster members in the magnitude range $r$=17-20, or
$-18.6<M_r<-15.6$ at the distance of A2199.  This range samples dwarf
galaxies in A2199 and therefore offers an excellent test of the
abundance of dwarf galaxies in dense environments.

We obtained optical spectroscopy of A2199 with MMT/Hectospec in 2007
July under marginal observing conditions.  Hectospec is a 300-fiber
multiobject spectrograph with a circular field of view of 1$^\circ$
diameter \citep{hectospec}.  We used the 270-line grating, yielding
6.2\AA~FWHM resolution.  

We observed A2199 with two configurations and obtained 479 secure
redshifts (out of 482 galaxies targeted).  We obtained 3 exposures of
600s for each configuration to facilitate cosmic ray removal.  Targets
in the first configuration included all galaxies in the magnitude
range 17$<r<$19.  Targets in the second configuration included
galaxies in the range 17$<r<$20, with rankings assigned to four
groups: 1) 17$<r<$19 galaxies on the photometric red sequence or
blueward [$g-r = -0.035(r-12)+1.0$], 2) 17$<r<$19 galaxies redward of
the red sequence, 3) 19$<r<$20 galaxies on or blueward of the red
sequence, and 4) 19$<r<$20 galaxies redward of the red sequence.  We
consider galaxies to lie on the red sequence or blueward if their
colors are no more than 0.1 mag redder than the red sequence (Figure
\ref{rs}).  We remove all galaxies with fiber magnitudes
$r_{fib}$$>$21 because they are unlikely to yield reliable redshifts
in short exposures with Hectospec.  We discuss the impact of this
selection in $\S \ref{lsbdiscuss}$.  Note that one galaxy with a Hectospec
redshift in A2199 (RA: 16h29m00.39s, DEC: +39:36:48.8 J2000) is
blended with a star in SDSS so that we do not have reliable photometry
for it.  As a rough estimate for the magnitude of the galaxy, we
subtract the flux for the star (determined from psfMag) from the
merged object to find $r\approx 19.6$.

The Hectospec field of view covers projected radii $R_P\leq1.11\Mpc$,
equivalent to $R_P=0.69 r_{200}\approx r_{500}$ for the parameters given in
\citet{cairnsi}, or $0.76 r_{200}$ for the parameters given in
\citet{cirsi}.  There are 32 galaxies in the Hectospec sample that 
have SDSS spectroscopy.  The mean velocity difference is
$-11.4\pm8.4~\kms$, and the scatter in the velocity differences is
$47.8~\kms$, slightly smaller than the mean uncertainty of $59.6~\kms$
calculated from the formal uncertainties.  Table \ref{a2199cz} lists
the coordinates and redshifts for the Hectospec data.  Columns 1 and 2
list the coordinates (J2000), Columns 3 and 4 list the heliocentric
velocity $cz$ and the corresponding uncertainty $\sigma_{cz}$, and
Column 5 lists the cross-correlation score R
\citep{km98}.

\begin{deluxetable}{lccccc}  
\tablecolumns{6}  
\tablewidth{0pc}  
\tablecaption{Spectroscopic Data for A2199\tablenotemark{a}\label{a2199cz}}  
\small
\tablehead{  
\colhead{}    RA & DEC & $cz$ & $\sigma _{cz}$ & R  \\
\colhead{}    (J2000) & (J2000) & ($\kms$)  & ($\kms$) & & \\
}
\startdata
 16:28:50.87 & 39:41:03.8 & 9074 & 37 & 08.05  \\
 16:28:46.96 & 39:44:59.8 & 10783 & 17 & 16.70  \\
 16:27:35.79 & 39:56:53.1 & 72585 & 35 & 12.91  \\
 16:28:50.56 & 40:00:46.0 & 83201 & 23 & 07.22  \\
 16:27:59.92 & 39:49:48.6 & 9331 & 56 & 04.72  \\
\enddata	     						  
\tablenotetext{a}{The complete version of this table is in the
 electronic edition of the Journal.  The printed edition contains only
 a sample.} 
\end{deluxetable}

To measure the luminosity function of the brighter galaxies in A2199,
we use redshifts for 207 galaxies measured either in SDSS or the
literature sources compiled in the NASA/IPAC Extragalactic Database
(NED)\footnote{http://nedwww.ipac.caltech.edu/}.  Many of these redshifts are from the
Cluster And Infall Region Nearby Survey (CAIRNS), which is complete to
$M_{K_s}$$\approx$-22.55 + 5 log $h_{70}\approx M^*+2$
\citep{rines02,cairnsii}, corresponding to $r$$\approx$16.  SDSS is
nominally complete to $r$=17.77 \citep{strauss02}, although fiber
collisions are more problematic for completeness in dense regions such
as nearby clusters.  
There are 12
galaxies in the range 16$<r<$17 that do not have redshifts in either
SDSS or NED.  Artificially including all these galaxies as members
does not change the LF significantly.

\subsection{Virgo}
We use recently released SDSS spectroscopy from Data Release 6
\citep[DR6;][]{dr6} to determine the luminosity function in the Virgo
cluster, the closest large galaxy cluster.  Virgo's proximity
\citep[$d\approx 17$~Mpc,][]{tonry01,mei07} allows the deepest possible
probes of the luminosity function.  However, Virgo is clearly
unrelaxed dynamically, as shown in the lumpiness of the galaxy
distribution \citep{binggeli85} and of the X-ray gas
\citep{bohringer94}.  

The SDSS DR6 data cover virtually the entire
sky within a projected radius of 1 Mpc from the central galaxy M87.
We focus our efforts on this region, although data are available in a
strip extending to much larger radius.  Galaxies within
1 Mpc of M87 are almost all contained within the main ``A'' cluster
\citep{binggeli85}.  The radial range covered is similar to that
covered in A2199 both in physical units and in overdensity: 1 Mpc in
Virgo is $\approx0.65 r_{200}$ \citep{mclaughlin99}.   

Many previous studies have
estimated the luminosity function in Virgo
\citep[e.g.,][]{binggeli85,impey88,phillipps98,trentham02,sabatini03},
but none have complete spectroscopy to the depth of SDSS.  Without
complete spectroscopy, previous investigations have relied on
either statistical methods of background subtraction or on alternative
membership indicators including morphology or surface brightness.


The proximity of the Virgo cluster presents challenges for
constructing a robust photometric catalog containing galaxies of
varying size, morphology, and surface brightness.  We detail the
construction of our new Virgo cluster catalog in the Appendix.

Figure \ref{virgors} shows the color-magnitude diagram of galaxies in
the Virgo cluster and in the background.  Colors are from
the fiber magnitudes \citep{dr6}.  Only four spectroscopically confirmed
members in the magnitude range $r$=13-16 redder than the red sequence
are Virgo members; all four lie on the red sequence or blueward if the
color is measured from the model magnitudes.  Nearly all Virgo members
with $r>16$ and redder than the red sequence are low surface
brightness galaxies and may have unreliable colors.  

\begin{figure}
\figurenum{2}
\label{virgors}
\plotone{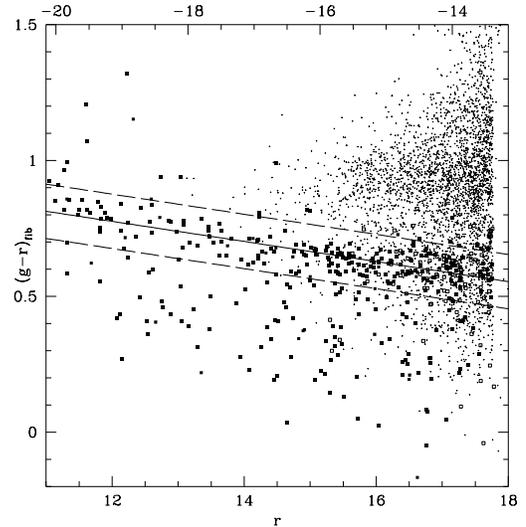}
\caption{Color-magnitude diagram for Virgo.  The red sequence 
is clearly visible (solid line), although it is distorted at the
bright end, possibly due to known problems with SDSS photometry of
bright galaxies.  Filled squares indicate spectroscopically confirmed
Virgo members, open squares indicate probable members lacking reliable
redshifts, and small dots indicate spectroscopically confirmed
background galaxies. Dashed lines indicate the color-magnitude cuts we
adopt for the red sequence of Virgo.  Note that again essentially all
galaxies redder than the red-sequence with spectra are background
galaxies. }
\end{figure}

\begin{figure}
\figurenum{3}
\plotone{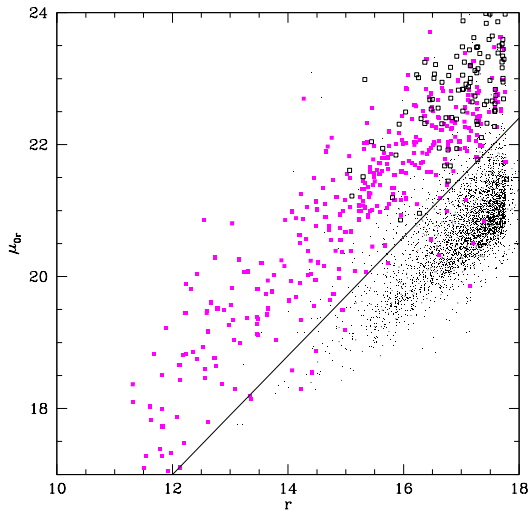}
\caption{\label{virgolsb} 
Central surface brightness versus apparent magnitude for galaxies
within 1 Mpc of M87.  Filled (open) squares indicate spectroscopically
confirmed (probable) Virgo members.  Small dots indicate background
galaxies.  The straight line indicates an approximate division between
Virgo members and background galaxies.  }
\end{figure}

Because surface brightness correlates with absolute magnitude, the
faintest Virgo galaxies in SDSS may be close to the surface brightness
limit of the survey.  \citet{blanton05} studied the completeness of
the SDSS pipeline using simulated images of galaxies with a wide range
of apparent magnitudes and surface brightnesses.  Figure
\ref{virgolsb} shows central surface brightness versus apparent magnitude for
galaxies within $R_p\leq 1 \mbox{Mpc}$ of M87.  We define the central
surface brightness as $\mu_{0r}=r_{Petro}+2.123$ to convert the fiber
magnitudes into mag arcsec$^{-2}$ (this definition assumes constant
surface brightness within the fiber).  Galaxies in the background of
Virgo tend to have higher surface brightness at a fixed apparent
magnitude \citep[][]{tolman,kurtz07}, but the loci of Virgo galaxies
and background galaxies overlap.  Galaxies from DR6 with $z\leq0.01$
but outside of Virgo show a similar distribution, indicating that this
difference is not due to photometric issues specific to the Virgo
cluster.

\epsscale{1}

The dramatically different distributions of magnitude versus surface
brightness for Virgo members and background galaxies strongly support
the use of surface brightness as a membership classification
\citep[e.g.,][]{binggeli85,conselice02,hilker03,mahdavi05,mieske07}.  
The SDSS spectra show the power of this classification.  Adopting
$\mu_{0r}=0.9r_{Petro}+6.2$ to separate the two populations (and
excluding 34 very bright galaxies with $r_{fib}<16$ that lie outside
Figure \ref{virgolsb}), 65.2\% of galaxies with lower surface
brightness are spectroscopically confirmed Virgo members.  Virgo
members comprise only 0.67\% of galaxies with higher surface
brightnesses.  This simple photometric cut removes 95.2\% of the
spectroscopically confirmed background galaxies.  

The power of this
technique shows that there is a tight correlation between absolute
magnitude and surface brightness for both cluster galaxies
\citep[e.g.,][]{andreon02b} and field galaxies
\citep[e.g.,][]{blanton05}.  This classification may be unusually
clean for the Virgo cluster due to the large deficit of galaxies in
the immediate background of Virgo \citep{ftaclas84}.  

\begin{figure}
\figurenum{4}
\plotone{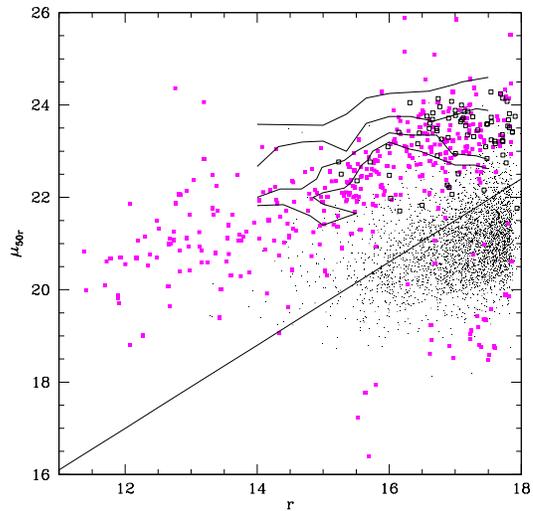}
\caption{\label{virgolsb2} 
Average surface brightness versus apparent magnitude for galaxies
within 1 Mpc of M87.   Filled (open) squares indicate spectroscopically
confirmed (probable) Virgo members.  Small dots indicate background
galaxies.  The straight line indicates an approximate
division between Virgo members and background galaxies. The four
curves indicate from top, 50\%, 75\%, 90\%, and 95\% completeness
contours for the SDSS imaging pipeline \citep{blanton05}.  
}
\end{figure}

Figure \ref{virgolsb2} shows average surface brightness within the
Petrosian half-light radius versus apparent magnitude.  This is the
definition of surface brightness used to construct the SDSS
spectroscopic target catalogs.  Figure \ref{virgolsb2} shows the
results of a completeness study of LSB galaxies performed by
\citet{blanton05}.  There are significant numbers of Virgo members in
the region where the SDSS spectroscopic target catalog begins to
become incomplete.  This incompleteness is mitigated by the inclusion
of large galaxies inserted by hand after the main target selection
\citep{blanton05} and by the procedures we follow here to identify
additional Virgo members below the spectroscopic target limits (see
Appendix).

\epsscale{1}

Figure \ref{virgolsb} provides a useful constraint on the abundance of
high surface brightness galaxies in clusters.  In particular, a new
class of high surface brightness galaxies was discovered in the Fornax
cluster \citep[][]{drinkwater99,hilker99}.  These galaxies, termed
ultracompact dwarf galaxies (UCDs), usually are unresolved in
ground-based imaging.  The typical luminosities of UCDs place them
between globular clusters and compact elliptical galaxies like M32.
Two groups have found UCDs in the Virgo cluster
\citep[][]{hasegan05,jones06}, but they appear to be a relatively
rare type of galaxy.  While many UCDs would be unresolved in
ground-based SDSS imaging, the SDSS data successfully recovers the
compact Virgo members VCC 1313 and VCC 1627 \citep{trentham02} as well
as a previously undiscovered UCD \citep[classified as a galaxy by the
SDSS photometric pipeline and described by][]{chilingarian07}.  

The SDSS spectroscopy demonstrates
conclusively that the Virgo cluster contains very few high surface
brightness galaxies that are resolved in ground-based imaging.
Estimating the total number of stellar-like UCDs in Virgo is
observationally expensive, requiring spectroscopy of all stellar-like
objects \citep{jones06} or HST imaging
\citep{hasegan05}.  Existing studies indicate that UCDs are not
sufficiently common to significantly affect the luminosity function of
Virgo cluster galaxies.

\section{Results}

We use the 479 Hectospec redshifts for A2199 along with 207 redshifts
from SDSS and the literature to determine the luminosity function.  Of
these 686 galaxies, 351 are members of A2199.  In Virgo, we find 484
definite or probable members (including 5 with $r$$\geq$17.77) out of
a total of 3971 galaxies within 1 Mpc of M87 and $r$$<$17.77.  The
luminosity functions for both clusters suggest relatively shallow
faint-end slopes, $\alpha=-1.13\pm0.07$ for A2199 and
$\alpha=-1.28\pm0.06$ for Virgo.

\subsection{A2199}

\subsubsection{Membership Fractions and Composition of Cluster Members}

Figure \ref{rs} shows that very few galaxies redward of the red
sequence are members of A2199.  Seven of the nine galaxies above our
nominal cutoff for red-sequence galaxies are $\leq$0.05 mag redder
than the cutoff (five are $\leq$0.02 mag redder), suggesting that our
cutoff might be too restrictive.  Inspection of the two remaining
galaxies reveals that their $g-r$ colors are likely overestimated due
to deblending problems in SDSS: the $g-r$ colors based on the SDSS
fiber magnitudes place both objects onto the red sequence.  Figure
\ref{virgors} shows a similar trend for the Virgo cluster; nearly all
galaxies redder than the red sequence are background galaxies.  The
exceptions are either deblending problems or low surface brightness
galaxies for which accurate colors are difficult to obtain ($\S 2.3$).
This result suggests that the red sequence is a remarkably well
defined limit for the intrinsic colors of cluster galaxies.  There are
no large populations of edge-on spiral galaxies or dusty starbursts
that have unusually red colors.

One striking aspect of Figures \ref{rs} and \ref{virgors} is that many
galaxies that lie on the photometric red sequence (and blueward) lie
well behind the cluster.

\begin{figure}
\figurenum{5}
\plotone{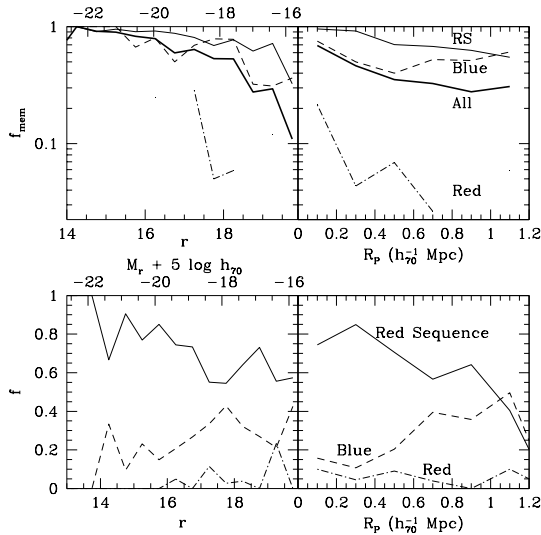}
\caption{\label{a2199frac} 
Top panels: Fraction of spectroscopically observed galaxies in A2199
that are cluster members as a function of apparent magnitude (left) or
projected radius (right).  The lines indicate the fractions for
galaxies on the red sequence (RS), redder than the red sequence
(``Red''; also shown by isolated points), and bluer than the red
sequence (``Blue'').  Bottom panels show the fraction of the cluster
population in each of these categories as a function of apparent
magnitude (left) or projected radius (right).  The blue fraction
increases with either increasing apparent magnitude or increasing
clustrocentric radius. }
\end{figure}

We quantify these trends in Figure \ref{a2199frac}.  We divide the
galaxies into three populations: red sequence galaxies within $\pm$0.1
mag of our assumed red sequence, and ``very red'' and ``blue'' galaxies for
galaxies outside this color range.  The upper left panel of Figure
\ref{a2199frac} shows the membership fraction as a function of
apparent magnitude for these three classes as well as for the total
population.  Using these membership fractions, we define the cluster
population by assuming that these membership fractions are constants
for each population.  

\epsscale{1}

The lower left panel of Figure \ref{a2199frac}
shows the fraction of the total cluster population in each of the
three classes.  Galaxies on the red sequence dominate the cluster
population at all magnitudes.  Because most of the member galaxies in
the ``very red'' population lie close to the cutoff, the fraction of
``very red'' cluster members is almost certainly overestimated.
Similarly, the membership fraction of the very red population should
be regarded as an upper limit.

The right-hand panels of Figure \ref{a2199frac} show these fractions
as functions of projected clustrocentric radius.  As expected, the
membership fractions generally decline with radius, and the fraction
of blue galaxies increases with radius \citep[and hence decreasing
density, e.g.,][]{1996ApJ...471..694A,balogh03,tanaka04,cairnsha}.
One surprising feature of the upper right panel of Figure
\ref{a2199frac} is that the membership fraction of blue galaxies does
not decline monotonically with radius but instead reaches a minimum
and then increases until it crosses the trend for red galaxies.

\subsubsection{The Luminosity Function of A2199}

\begin{figure}
\figurenum{6}
\label{lf}
\plotone{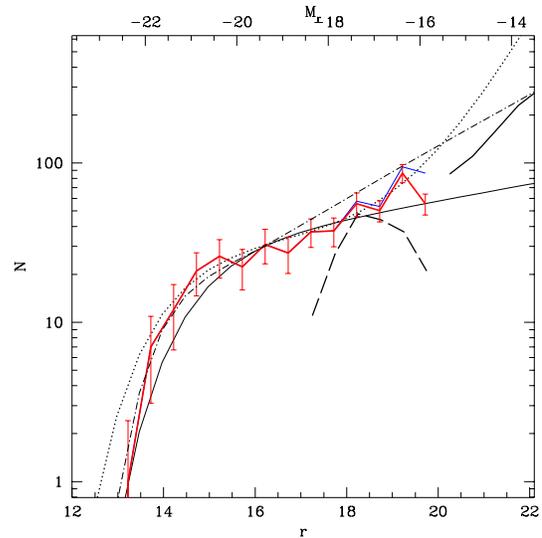}
\caption{Luminosity function of A2199 determined using spectra 
from MMT/Hectospec and SDSS DR6 (thick red solid line).  Errorbars
indicate Poissionian uncertainties.  The blue line shows the results
of a correction for low surface brightness galaxies ($\S 4.1.5$).  The
dashed line at 17$<$$r$$<$20 shows spectroscopically confirmed members
from our Hectospec data.  The rising dotted line indicates the cluster
luminosity function of
\citet{popesso06b} and the dash-dotted line indicates the field LF of
\citet{blanton05}.   The thick solid line at faint
magnitudes shows an extrapolation of the LF 
assuming membership fractions for the color bins of
$f_{RS}=f_{blue}=0.3$ and $f_{red}=0$ (Figure \ref{a2199frac}).  The
short-dashed lines at faint magnitudes show the extrapolated LF
assuming (upper) all LSB galaxies are members or (lower) that
$f_{mem}$ is the same as in Virgo at the comparable absolute magnitude
(Figure \ref{virgofrac}). }
\end{figure}

Based on Figures \ref{rs} and \ref{a2199frac}, we estimate the total
luminosity function in 0.5 mag bins by applying the membership
fraction of spectroscopically observed galaxies to the three color
populations.  Figure \ref{lf} shows the luminosity function of A2199.
The upper solid line shows the counts of galaxies in 0.5 mag bins when
restricted to the ``red sequence'' and ``blue'' cuts defined above.
The raw counts of galaxies on or bluer than the red sequence show a
significant upturn at $M_r$=-17, similar to the upturn seen by
\citet{popesso06b}.  

The dashed line at 17$<$$r$$<$20 shows the counts
of spectroscopic members, where members have 7000$<cz<$11,000~$\kms$
\citep[][]{rines02}.  Note that the contrast in redshift space between
cluster members and background galaxies is large; the exact
velocity limits are not a significant source of uncertainty
\citep{rines02,cirsi}.  The thick solid line in Figure \ref{lf} shows
the resulting LF.

The LF of A2199 can be well fit by a function with the form proposed
by \citet{schechter76}.  This function has the form
\beqn
\frac{dn}{dM} \Bigg{\vert}_M \propto 10^{0.4(1+\alpha)(M_*-M)} {\rm exp} [-
10^{0.4(M_*-M)}]
\eeqn
where $M^*$ is the characteristic magnitude of the LF and $\alpha$ is
the faint-end slope.  We fit the A2199 LF by minimizing $\chi^2$ and
excluding the BCG.  We find $M^*=-21.11^{+0.21}_{-0.25}$ and
$\alpha=-1.13^{+0.07}_{-0.06}$ (thin solid line in Figure \ref{lf}).
Both parameters are similar to those found for field galaxies in SDSS
\citep{blanton03} and the Century Survey \citep{csccdlfn}. 
\citet{blanton05} investigate the field LF to fainter absolute
magnitudes with SDSS and find that it is consistent with steeper
faint-end slopes of $\alpha\approx-1.4$.  The LF of A2199 is therefore
similar to or perhaps shallower than the field LF.  We discuss
systematic uncertainties in the LF of A2199 below.  An independent,
shallower ($M_V<-17$) study of the A2199 LF found
$\alpha=-1.12\pm0.06$ \citep{andreon07}, in excellent agreement with
our result.

Figure \ref{lf} shows an extrapolation of the A2199 LF to fainter
magnitudes.  The thick solid line at faint magnitudes shows the LF
inferred by assuming that the membership fraction $f_{mem}$ for the
three color bins remains constant at fainter apparent magnitudes,
using the faintest bin with spectroscopy to estimate these values of
$f_{mem}$.  Because $f_{mem}$ decreases steeply with apparent
magnitude, this extrapolation is an approximate upper limit on the LF.
We make two additional estimates of the extrapolated LF in A2199.  The
first assumes that all galaxies with $r_{fib}>21$ in the ``blue'' and
``red-sequence'' bins are cluster members.  This extrapolated LF
yields a more conservative upper limit to the A2199 LF; this
extrapolation contains a factor of 2 fewer galaxies than the Popesso
LF at $M_r\approx-14$.  The final estimate of the extrapolated LF
assumes that $f_{mem}$ at a given absolute magnitude is the same as in
the Virgo cluster (Figure \ref{virgofrac}; $\S 3.2.1$).
Interestingly, the values of $f_{mem}$ for Virgo and A2199 agree at
the absolute magnitudes where they overlap, perhaps because the data
extend to similar fractions of $r_{200}$ for the two clusters.  This
Virgo-based extrapolation of the LF is consistent with extending the
best-fit Schechter function.  Thus, the range of lines at faint
magnitudes in Figure \ref{lf} encompasses the full range of reasonable
extrapolations of the LF in A2199.

\subsection{Virgo}

\subsubsection{Membership Fractions and Composition of Cluster Members}

\begin{figure}
\figurenum{7}
\plotone{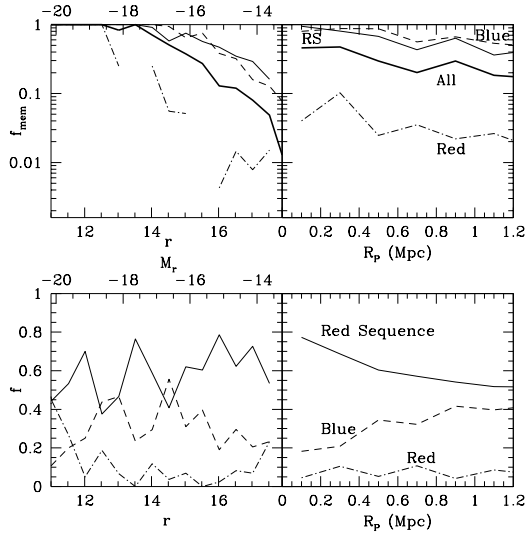}
\caption{\label{virgofrac} 
Top panels: Fraction of spectroscopically observed galaxies in Virgo
that are cluster members as a function of apparent magnitude (left) or
projected radius (right).  The lines indicate the fractions for
galaxies on the red sequence (RS), redder than the red sequence
(``Red''), and bluer than the red sequence (``Blue'').  Bottom panels
show the fraction of the cluster population in each of these
categories as a function of apparent magnitude (left) or projected
radius (right).  The blue fraction increases with either increasing
apparent magnitude or increasing clustrocentric radius. }
\end{figure}

Figure \ref{virgofrac} shows the membership fraction of the three
galaxy populations (red sequence, very red, and blue) in the Virgo
cluster.  The large deficit of galaxies in the immediate background of
Virgo \citep{ftaclas84} means that the membership cuts in redshift
space are robust.  The scatter in $g-r$ color increases with
increasing apparent magnitude (Figure \ref{virgors}), although part of
this increased scatter reflects less reliable photometry for low
surface brightness galaxies (Figure \ref{virgolsb}).
\citet{conselice03} find a similar increase in scatter of colors of
fainter galaxies in the Perseus cluster; they interpret this result as
evidence that faint galaxies have more complex star formation
histories than bright galaxies.  To avoid the larger photometric
uncertainties in these faint galaxies, we show the radial trends in
Figure \ref{virgofrac} only for galaxies with $r<16$ so that these can
be compared directly with A2199.  

\epsscale{1}

Many of the
trends apparent in Figure \ref{a2199frac} for A2199 are also apparent
for Virgo.  Membership fractions decline with radius for all
populations, and $f_{mem}\lesssim 0.1$ for very red galaxies.  Virgo
members are dominated by red-sequence galaxies, and the blue fraction
increases (weakly) with radius.  Again, the membership fractions of
blue galaxies and red sequence galaxies are quite similar at large
projected radii.


\subsubsection{The Luminosity Function of Virgo}

\begin{figure}
\figurenum{8}
\label{virgolf}
\plotone{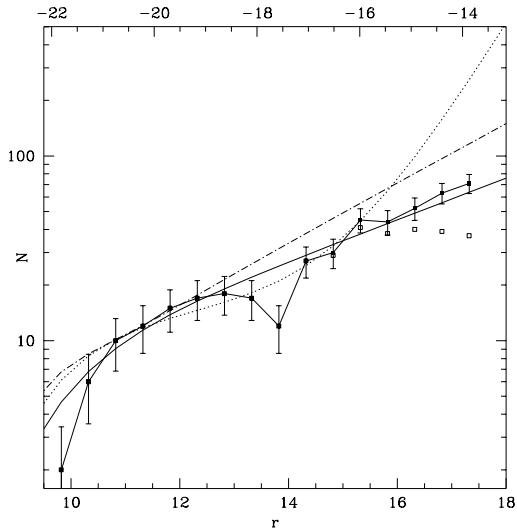}
\caption{Luminosity function of the Virgo cluster within 1 Mpc 
of M87 determined using spectra from SDSS DR6 (squares).  Errorbars
indicate Poissionian uncertainties.  The thin solid line shows the
best-fit Schechter function.  Open squares show the ``minimal''
luminosity function including only spectroscopically confirmed
background galaxies.  The rising dotted line indicates the cluster
luminosity function of \citet{popesso06b} and the dot-dashed
line shows the field LF from SDSS \citep{blanton05}.  }
\end{figure}

Figure \ref{virgolf} shows the luminosity function of the Virgo
cluster within 1 Mpc of M87.  The bright end of the LF is poorly
constrained due to the difficulty of bright galaxy photometry in SDSS
\citep{dr6}.  Open squares show the ``minimal'' luminosity function
including only spectroscopically confirmed background galaxies.
Figure \ref{virgofrac} shows the membership fraction $f_{mem}$ as a
function of apparent magnitude.  Note that $f_{mem}$ decreases from
unity at bright magnitudes to $\sim$0.03 at $M_r\approx-14$.
Therefore, any attempt to measure the cluster LF at such faint
magnitudes using statistical background subtraction requires a
determination of the background population to per cent level accuracy
{\it and} uniformity.  

The number of background galaxies exceeds the number of Virgo galaxies
at $r\approx15$ \citep[first suggested by][and confirmed
here]{shapleyames}, or $M_r\approx-16.5$, close to the absolute
magnitude where the cluster LF from the stacked SDSS analysis
indicates an upturn in the cluster LF \citep{popesso05a,popesso06b}.
There is a pronounced dip in the Virgo LF at $r\approx14$.  Because
galaxies at these magnitudes may have inaccurate photometry from the
SDSS pipeline \citep{dr6}, we caution that this dip may be an artifact
of the pipeline photometry.

The Virgo cluster LF is fit by a faint-end slope of
$\alpha=-1.28\pm0.06$ (statistical).  We hold $M_r^*=-21.32$ fixed at
the value in the field \citep{blanton03} due to the problem with
bright galaxy photometry in SDSS \citep{dr6}.  The faint-end slope
increases by less than $1\sigma$ ($\alpha=-1.24$) if we fit to the
``minimal'' luminosity function.

For comparison, we show the composite cluster LF
from \citet{popesso06b}, which has a steep upturn at faint magnitudes.
We find no such upturn in the Virgo cluster.  Figure
\ref{virgolf} also shows the field LF determined from SDSS data
\citep{blanton05}.  This field LF is slightly steeper than the Virgo
LF, although the Virgo LF is consistent with the faint-end slope of
$\alpha=-1.32$ found for the raw counts in \citet[][i.e., with no
correction for missing LSB galaxies]{blanton05}.  The faint-end slope
we measure for Virgo is significantly steeper than that in A2199.
Because of the different ranges of absolute magnitude sampled, it
would be premature to conclude that there is true variation in the
faint-end slope among clusters.

One remarkable aspect of our new determination of the Virgo LF is that
it lies very close to the one determined by \citet{sandage85}, who
found a faint-end slope of $\alpha\approx-1.25$ from the raw counts
and $\alpha=-1.30$ after correcting for incompleteness at the surface
brightness limit of the survey.  Another interesting feature of our
determination of the Virgo LF is that it requires assumptions about
the membership of low surface brightness galaxies. \citet{sandage85}
assumed a relation between absolute magnitude and surface brightness
(based on galaxies in the RSA catalog) to conclude that only low
surface brightness galaxies are Virgo members.  The extensive SDSS
spectroscopy demonstrates that the vast majority of high surface
brightness galaxies are indeed in the background, but the low surface
brightnesses of the presumed Virgo members prevents them from having
well-measured redshifts even with SDSS spectroscopy.

\section{Discussion}

We separate our discussion of the cluster luminosity function into two
parts.  First, we discuss possible systematic effects in the detection
and photometry of galaxies that may affect the LF.  Second, we discuss
the astrophysical implications of the cluster LF.

\subsection{Systematic Effects in Determining the Cluster LF}

\subsubsection{Treatment of Red Galaxies}

Figure \ref{rs} shows that there are essentially no cluster galaxies
significantly redward of the red sequence.  At least some of the
apparent discrepancy between the LF in A2199 and claims of steeper
slopes can be explained by the treatment of red galaxies, in
particular those redward of the red sequence.  Some studies contain no
color information \citep{sandage85,phillipps98}.  They are thus unable
to identify background galaxies by using the red sequence.

\citet{jenkins07} use SDSS colors to select candidate members in 
Coma from an IRAC
object catalog using a generous color cut of $g-r$$<$2, despite the
fact that there are no spectroscopically confirmed members redward
of $g-r$=1.0 (their Figure 11).  They then use the optical
spectroscopy of \citet{mobasher03} to determine the fraction of
galaxies that are Coma members in each IRAC Ch1 magnitude bin.  If the
spectroscopic target selection were independent of $r-\mbox{Ch1}$
color, this procedure would yield an accurate estimate of the
membership fraction in each IRAC bin.  However, the spectroscopic
targets are strongly biased to lie blueward of the large
population of faint red galaxies (their Figure 14).  Thus, this
membership fraction is biased high and the resulting LF is probably
artificially steep.

A caveat to the conclusion that nearly all ``very red'' galaxies are
background galaxies is the larger scatter in the color-magnitude
diagram at faint magnitudes \citep[$\S 3.2.1$;][]{conselice03}.  This
increased scatter may cause very faint cluster members to be scattered
redward of the red sequence.

\subsubsection{Constraints on the LF From Gravitational Lensing}

The deep gravitational potentials of clusters produce gravitational
lensing of background galaxies.  Lensing distortions are usually
measured from systematic distortions in the shapes of faint galaxies
which lie behind the cluster. 
A fascinating application of gravitational lensing to the LF problem
was implemented by \citet{medezinski07}.  Using deep HST/ACS
photometry of A1689, they showed that the weak lensing signal depends
on the location of faint galaxies relative to the cluster's red
sequence.  The shear signal is stronger for galaxies redward of the
red sequence than it is for galaxies on the red sequence.  The
diminution of the shear signal for galaxies on the red sequence is
directly correlated with the fraction of galaxies that are cluster
members and therefore unlensed.  The amplitude of the diminution in a
given magnitude bin is then a measurement of the membership fraction
in that bin.  The LF estimated with this procedure has a significantly
shallower slope ($\alpha=-1.05\pm0.1$) than many estimates based on
background subtraction.  

Figures \ref{rs} and \ref{virgors} support this approach.  Few
galaxies redward of the red sequence are cluster members.  Thus, the
lensing signal measured from these galaxies suffers little dilution
from cluster members.

\subsubsection{Photometry Around Bright Galaxies \label{pieces}}

As part of a weak lensing analysis of systems in SDSS,
\citet{mandelbaum05} found that SDSS catalogs contain a deficit of
objects near bright galaxies.  They conclude that this deficit is
produced by overestimation of the sky background because of the
stellar haloes of the bright galaxies.  Because bright galaxies are
significantly more common in fields containing galaxy clusters than in
offset fields, there should be {\it fewer} objects in cluster fields.
This deficit could lead to an apparent deficit of faint galaxies in
cluster fields and hence potentially mask a steep LF.

We noticed a related effect which may counteract this deficit during
the visual inspection of Virgo cluster galaxies.  Many large galaxies,
especially those with low surface brightnesses, have small pieces 
detected as separate objects.  These detections have little
effect on the photometry of the ``parent'' galaxies because they are
usually $>$3 mag fainter than the parent galaxy.  However, these
pieces are included in the SDSS Galaxy tables.  If these pieces of
galaxies are not removed, they could produce an artificial excess of
faint galaxies in cluster fields.

\subsubsection{Determining Membership from Photometry and Morphology}

The VCC assigned membership via the photometric and morphological
properties of galaxies.  It is quite striking that our Virgo LF
reproduces their results even after the extensive spectroscopy of
SDSS.  We saw in $\S 2.3$ (Figure \ref{virgolsb}) that Virgo galaxies
are cleanly separated from background galaxies in the distribution of
apparent magnitude versus central surface brightness.  The relatively
tight relation between absolute magnitude and surface brightness
assumed by \citet{sandage85} enabled them to efficiently select Virgo
members and determine the Virgo LF accurately.   The SDSS
spectroscopy confirms that most of the galaxies omitted from the VCC
due to photometric properties are indeed background galaxies.  At
faint magnitudes, the SDSS spectroscopy is not sufficient to confirm
the Virgo membership of many of the ``Possible'' members from the VCC
(due to their low surface brightness).

The study of the NGC 5846 group by \citet{mahdavi05} uses a similar
approach to the one used here: using spectroscopy of a subsample to
estimate the fraction of galaxies identified as probable members by
their photometric properties.  They find that all galaxies they
classify as ``probable'' members are in fact members, while about half
of the ``possible'' members are true members.

\subsubsection{Low Surface Brightness Galaxies \label{lsbdiscuss}}

One longstanding problem in determining the LF is the low surface
brightness of dwarf galaxies \citep[e.g.,][]{impey88}.  Indeed, some
investigators have used the correlation between magnitude and surface
brightness as a way of identifying likely cluster members
\citep{conselice02,mieske07}.  However, low surface brightness
galaxies can be difficult to detect, especially in shallow photometric
surveys such as SDSS \citep[see discussion in][]{blanton05} and they
are challenging objects for accurate spectroscopy.

\begin{figure}
\figurenum{9}
\plotone{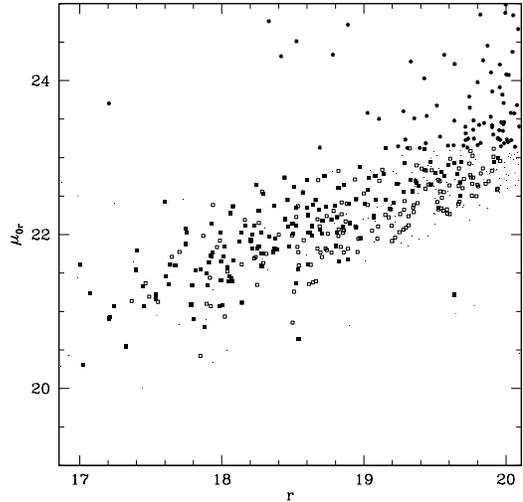}
\caption{\label{a2199lsb} 
Central surface brightness versus apparent magnitude for galaxies in
A2199.  Cluster members are large squares and background galaxies are
small dots.  Galaxies from DR6 with $z<0.01$ are shown as small
crosses.  Stars indicate low surface brightness objects with
$r_{fib}>21$ which were excluded from spectroscopic targeting. }
\end{figure}

We show the distribution of surface brightness as a function of
apparent magnitude for A2199 in Figure \ref{a2199lsb}.  At the faintest
magnitudes probed ($r=19-20$), there are several galaxies excluded
from spectroscopic targeting due to their low surface brightnesses,
fiber magnitudes $r_{fib}\geq 21$.  These low surface brightness
galaxies lie on both the red sequence and on the locus of surface
brightness and apparent magnitude traced by cluster members (Figure
\ref{a2199lsb}).  It therefore seems likely that many of these low
surface brightness galaxies are members of A2199.  Including these
galaxies steepens the LF slightly, and this systematic uncertainty is
large enough that the LF could be consistent with the LF of
\citet{popesso06b}.  Our analysis in $\S 3.1.2$ suggests that the
faint end of the A2199 LF is likely intermediate between the slope of
$\alpha=-1.13$ found for our spectroscopic sample and $\alpha=-1.4$
for field galaxies \citep{blanton05}.  It is quite possible that
the difference in the measured values of $\alpha$ for A2199 and Virgo
is created by missing dwarf galaxies in A2199, either from the SDSS
photometric pipeline or our spectroscopic cutoff of $r_{fib}<21$. 

\epsscale{1}

It is curious to note that the distinction between A2199 members and
background galaxies (Figure \ref{a2199lsb}) is not nearly as clean as
in Virgo (Figure \ref{virgolsb}).  This difference probably indicates
that there is lower contrast between A2199 and ``near-background''
galaxies (those within $\Delta z\approx 0.01$) than in the Virgo
cluster.

In the Virgo cluster, there have been many searches for low surface
brightness galaxies missing from the VCC
\citep[e.g.,][]{impey88,trentham02b,sabatini03}, including studies
with deep CCD imaging using the Subaru 8-m telescope.  Because these
studies have failed to reveal a much larger population of low surface
brightness galaxies, \citet{trentham02} conclude: ``the major concern
is now that the sample may be missing a sample of {\it high} surface
brightness galaxies that we have culled from the sample because we
think that they are background galaxies.''  With the extensive
spectroscopy available in SDSS DR6, the lingering concern over high
surface brightness galaxies can be laid to rest.

Unlike A2199, the difference between the Virgo LF and that of
\citet{popesso06b} cannot be explained by  missing LSB
galaxies.  The difference between the two LFs becomes significant at
$r\approx 15$ and is greater than a factor of 2 at $r\approx 16$.  At
these bright magnitudes, the incompleteness due to surface brightness
is not significant (Figure \ref{virgolsb2}).  Also, galaxies in Virgo
missed by SDSS due to surface brightness would be missed by SDSS in
the number counts used by \citet{popesso06b}.  

Our efforts to identify additional LSB galaxies in Virgo (see
Appendix) turned up a number of new galaxies, although these
additional galaxies have little effect on the LF.  The LF from the
Virgo catalog prior to these additions (essentially the spectroscopic
catalog with the addition of bright galaxies and LSB spectroscopic
targets with failed or no spectra) has a best-fit faint end slope of
$\alpha=-1.24$, or less than 1$\sigma$ different from our final
estimate.
The issue of low surface brightness galaxies remains a serious
systematic uncertainty fainter than $M_r\approx-14$ (beyond the limit
of the measurements presented here).  A systematic search for LSB
galaxies in the SDSS imaging but below the detection threshold of the
photometric pipeline would be very instructive, but such a search lies
beyond the scope of this work.  

\subsection{Implications of the Cluster LF}

An accurate determination of the LF in different environments is a
powerful constraint on models of galaxy formation and evolution.  When
previous work indicated a relative excess of dwarf galaxies in
clusters, attempts were made to explain this environmental dependence
due to the varying effects of reionization in different environments
\citep{tully02,benson03}.

\subsubsection{Radial Dependence of the LF}

One aspect of recent claims about the faint-end slope of the LF in
clusters is difficult to explain with the effects described in $\S
\ref{pieces}$. Several studies have claimed that the faint-end slope is
steeper at large clustrocentric radii and shallower in the cluster
core \citep[$R_p\sim300\kpc$; e.g.,][]{popesso06b,barkhouse07}.  The
potential excess from pieces of galaxies described in $\S
\ref{pieces}$ would naturally produce a faint-end slope which
increased continuously towards cluster centers: the density of bright
galaxies generally increases towards cluster centers.  The number of
galaxy pieces should do the same.  The shallower faint-end slope in
cluster centers suggested by \citet{popesso06b} and
\citet{barkhouse07} is therefore difficult to explain with this
potential systematic bias.

Some recent studies claim that LF fits also yield brighter $M^*$ at
larger clustrocentric radii.  The direction of the changes in both
parameters is along the well-known degeneracy between $\alpha$ and
$M^*$ in Schechter function fits.  There is, however, an important
systematic effect in these studies that is often ignored.  The BCGs
are usually explicitly excluded from the LF fits.  Because BCGs tend
to lie near cluster centers, this procedure acts to produce an
apparent brightening of $M^*$ with radius.

\begin{figure*}
\figurenum{10}
\label{virgolfr}
\plottwo{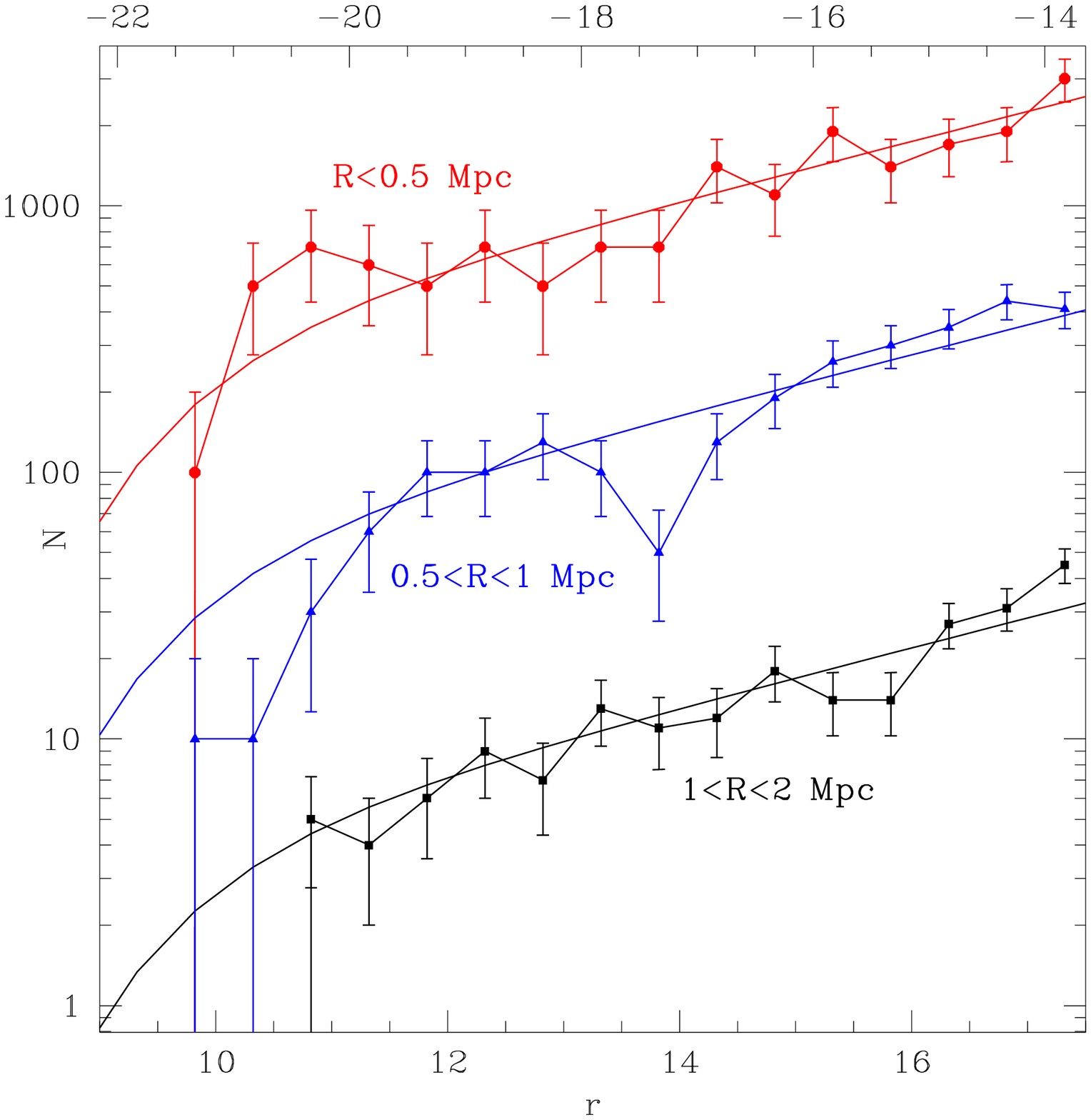}{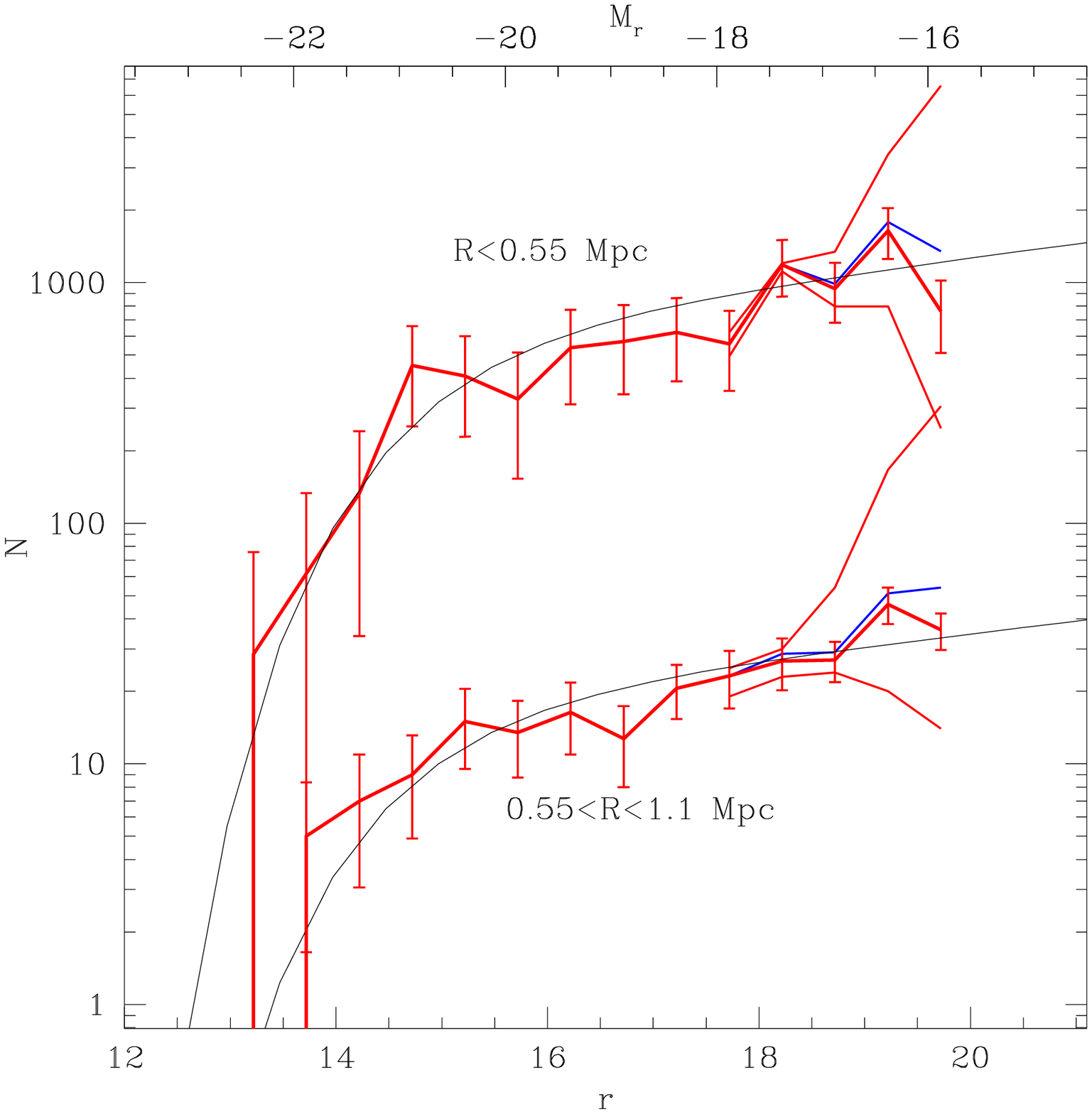} 
\caption{Cluster LF versus projected radius in (a) Virgo and 
(b) A2199.  The radial bins in Virgo are $R_P<0.5$ Mpc, 0.5-1.0 Mpc,
and 1.0-2.0 Mpc, or approximately $R_P<0.33$, 0.33-0.66, 0.66-1.3
$r_{200}$ \citep{mclaughlin99}.  Note that the outermost bin has only
partial spectroscopic coverage. The radial bins in A2199 are
$R_P<0.55$ Mpc and 0.55-1.1 Mpc, or approximately $R_P<0.3$$r_{200}$
and 0.33-0.66 $r_{200}$ \citep{cirsi}.  
}
\end{figure*}

Figure \ref{virgolfr} shows the LF as a function of radius in Virgo
and A2199.  We again show the LFs of
\citet{popesso06b,blanton05,trentham02b} for comparison (with
arbitrary normalization).  The LF does not have a strong dependence on
clustrocentric radius, counter to claims by \citet{popesso06b} and
\citep{barkhouse07}.  

There is some evidence that $M^*$ is brighter at
smaller radii in Virgo, but A2199 shows the opposite trend.  The trend
of brighter $M^*$ at smaller radii (and hence denser environments) is
expected due to the tendency of very luminous galaxies to inhabit the
densest environments \citep[e.g.,][]{hogg03,zehavi05}.  
Note that the dip in the Virgo LF at $r\approx14$
appears to be confined to the annulus 0.5$<R_P<$1.0 Mpc.  
A more
detailed treatment of the photometry of all Virgo members would be
required to test the significance of this dip.

\subsubsection{Total Cluster Luminosity: Intracluster Light and the LF}

One motivation for determining the LF in clusters is to assess the
total amount of starlight in a given cluster.  Recent studies of
intracluster light in nearby clusters have shown that it contributes a
significant fraction of the total cluster light
\citep[e.g.,][]{gerhard07}.

If we adopt the faint-end slope from Virgo, the contribution to total
galaxy light from galaxies fainter than $M_r^*+2$ ($M_r^*+3$) is
27.4$\pm$3.0\% (14.6$\pm$2.5\%).  Thus, the uncertainty in the total
cluster starlight due to faint galaxies is significantly smaller than
the uncertainty due to intracluster light, which may contribute 5-50\%
of the total stars in a cluster
\citep[e.g.,][]{mihos05,krick07,gonzalez07}.

Future studies of intracluster light which extend to very low surface
brightness will be very useful for quantifying the number and
luminosity of LSB galaxies missed in existing surveys.  Similarly,
future surveys of intracluster planetary nebulae and other tracers
should provide upper limits on the fraction of galaxian light in
clusters contributed by LSB galaxies below the detection limits of
galaxy surveys.

\section{Conclusions}

Determining the faint end of the luminosity function in clusters has
remained an unresolved problem for many years.  We report a new
estimate of the luminosity function in A2199 from MMT/Hectospec
spectroscopy, and of the Virgo cluster from SDSS DR6 data.  Both LFs
extends to fainter absolute magnitudes than most previous
determinations of the LF in massive clusters.  The LF closely follows
a Schechter function to $M_r\approx-15\approx M^*+6$ in A2199 and to
$M_r\approx-13\approx M^*+8$ in Virgo.  There are no large populations
of high surface brightness galaxies or galaxies redder than the red
sequence that contribute significantly to the LF in either cluster.

In A2199, we find that essentially no galaxies redward of the
photometric red sequence are cluster members.  However, the red
sequence itself contains a significant number of background galaxies.

We find no evidence of an upturn in the A2199 LF at faint magnitudes
as claimed by some recent studies
\citep[][]{popesso06b,jenkins07,milne07,adami07,yamanoi07,barkhouse07}, 
although the range of absolute magnitudes precludes a conclusive
result.  A simple extrapolation of the A2199 LF using the membership
fractions at the spectroscopic limit provides an approximate upper
limit of the LF at fainter magnitudes; this extrapolation suggests
that the faint end slope is probably comparable to that of Virgo.

In the Virgo cluster, we find an LF consistent with a moderate
faint-end slope ($\alpha=-1.28\pm0.06$).  The Virgo LF extends much
fainter than the A2199 LF, and we conclusively demonstrate that the
Virgo LF is inconsistent with the steeper LFs found by
\citet{popesso06b}.  

The discrepancy between the LFs may be due to
systematic uncertainties in statistical background subtraction, and we
discuss some possibilities.  Other estimates of the LF using deep
spectroscopy find slopes similar to ours
\citep{mobasher03,christlein03,mahdavi05}.  Recent estimates of the LF
in Fornax using surface brightness as a membership classification
\citep{hilker03} or surface brightness fluctuations to estimate
distances \citep{mieske07} indicate a shallow LF similar to the Virgo
SDSS LF.  Similarly, a clever application of gravitational lensing in
A1689 by \citet{medezinski07} suggests a LF consistent with those we
find in Virgo and A2199.

Low surface brightness galaxies remain problematic for determining
cluster LFs.  Their low surface brightnesses prohibit reliable
redshift estimates using SDSS DR6 spectroscopy.  Perhaps the most
significant impact of the SDSS spectra is to demonstrate conclusively
that higher surface brightness galaxies are virtually all in the
background of Virgo.  A simple division in apparent magnitude versus
surface brightness is a surprisingly powerful membership
classification (Figure \ref{virgolsb}, $\S 2.3$).  Careful inspection
of SDSS galaxies failing the spectroscopic target selection criteria
reveals many low surface brightness galaxies that are likely Virgo
members.  It is somewhat ironic that almost all $r<17.5$ galaxies
within 1 Mpc of M87 without well-measured redshifts are the LSB
galaxies most likely to be Virgo members.  
We list these galaxies in the Appendix as an aid for future efforts to obtain spectroscopy of these galaxies.

The spectroscopically determined LFs we find for the A2199 and Virgo
clusters are similar to the field LF, an important result for models
of galaxy formation.  Future studies of the LF in more clusters and at
larger clustrocentric radii will constrain cluster-to-cluster
variations in the LF and any radial dependence.  The fraction of
galaxies belonging to a cluster decreases dramatically with both
increasing magnitude and increasing projected radius.  Even at the
relatively small radii we probe here, we show that statistical
background subtraction is problematic due to the high precision
required.  We therefore recommend that future investigations of the LF
in clusters avoid statistical background subtraction and instead
identify member galaxies via spectroscopy or photometric information
such as colors or surface brightness fluctuations.  

\acknowledgements

We thank Andres Jordan, Warren Brown, and Michael Kurtz for helpful
 discussions and suggestions.  We thank Susan Tokarz for assistance
 with the reduction of Hectospec data.  Observations reported here
 were obtained at the MMT Observatory, a joint facility of the
 Smithsonian Institution and the University of Arizona.  This research
 has made use of the NASA/IPAC Extragalactic Database (NED) which is
 operated by the Jet Propulsion Laboratory, California Institute of
 Technology, under contract with the National Aeronautics and Space
 Administration.  Funding for the Sloan Digital Sky Survey (SDSS) has
 been provided by the Alfred P. Sloan Foundation, the Participating
 Institutions, the National Aeronautics and Space Administration, the
 National Science Foundation, the U.S. Department of Energy, the
 Japanese Monbukagakusho, and the Max Planck Society. The SDSS Web
 site is http://www.sdss.org/.

{\it Facilities:} \facility{MMT (Hectospec)}

%

\appendix

\section{Construction of the SDSS Virgo Cluster Catalog}

Here we describe our methods for analyzing the SDSS DR6 data and
identifying spectrocopically confirmed members of the Virgo cluster as
well as several additional galaxies that are probable members.  We
begin our analysis with a sample drawn from the ``Galaxy'' view of
SDSS DR6 from the CAS
server.\footnote{http://casjobs.sdss.org/CasJobs/} This database
contains basic photometric and spectroscopic parameters for galaxies
in the SDSS spectroscopic survey.  We address the issue of galaxies
without spectra below.

One disadvantage of the proximity of Virgo is that the largest
galaxies in Virgo are often ``shredded'' by the SDSS pipeline
processing, a problem discussed in detail in the context of measuring
the field LF \cite[][]{blanton05}.  To assess the importance of this
``shredding'' we visually inspect images of all 408 galaxies
classified as Virgo members according to the cut: $|\Delta cz|\leq
2000~\kms$, $R_p\leq1\mbox{Mpc}$ from M87 and spectral class equal to
``Galaxy''.  We edit the photometric catalog as follows: if a SDSS
galaxy is identified as being a part of a larger galaxy, we eliminate
the ``galaxy part'' from the catalog.  We find 52 ``galaxy parts'' in
the catalog.  If the larger galaxy is not included in the catalog, we
add it manually.  If no photometric object within the photometric
outline of the target galaxy is less than 3 mag (using model
magnitudes) fainter than the primary galaxy, we classify the galaxy
photometry as ``clean''.  We classify galaxies as ``clean'' if the
secondary component is clearly a separate object, i.e., an unrelated
star or galaxy.  We also search around the member galaxies for
unresolved galaxies.  There are very few of these, indicating that the
standard SDSS pipeline is more likely to split a galaxy into multiple
pieces than to merge multiple galaxies into one detection.  These
photometric splits very rarely have a large effect on the estimated
magnitude of the galaxy.  We explore the effects of adding back the
flux from secondary components versus ignoring the additional flux and
find that the effect on the LF is minimal.

We also visually inspect the 888 objects with spectra not classified
as ``Galaxy'' but satisfying the velocity criterion (thus including
all Milky Way stars) to find any misclassified galaxies.  Several
galaxies have their spectra classified as ``Star''.  Many of these
galaxy spectra have prominent Balmer features and very low redshifts
consistent with Galactic objects.  Several galaxies with poor spectra
are classified as ``Unknown''.  Many of these galaxies have low surface
brightness, and most of them are members of the Virgo Cluster
Catalog \citep[VCC,][]{binggeli85}.  Although the redshifts for many
of these low surface brightness galaxies are unreliable, they are
likely to be members of the Virgo cluster.  We compare SDSS to the VCC
below.

Similarly, we visually inspect all objects with spectra classified as
``Galaxy'' outside the redshift cuts if they have low-confidence
redshifts (zConf$<$0.9).  We again find many LSB galaxies with low
signal-to-noise spectra and incorrect redshifts.  Because most of
these galaxies are likely Virgo members, we include these in the
catalog of Virgo members.  

We match targets selected from the SDSS DR6 photometric ``Galaxy''
table to those with spectroscopy.  There are many more photometric
objects than spectroscopic objects brighter than $r_{Petro}=17.77$,
the limit of the spectroscopic survey for Main Sample galaxies.  We
inspected several hundred of these objects with the Image List tool
and confirm that they are nearly all misclassified stars or pieces of
stars (e.g., diffraction spikes or double stars).  The remaining
objects are mostly Main Sample target galaxies without spectra or
pieces of larger galaxies such as HII regions.  There are only a
handful of galaxies that have been missed, usually due to problems
with the photometry (e.g., poor deblending of nearby stars or
galaxies).  In particular, we do not find a large number of low
surface brightness galaxies that could have been excluded from the
Main Sample target selection.  We conclude that the Main Galaxy target
sample is an essentially complete sample of galaxies from the SDSS
database.

We use Petrosian magnitudes for Virgo galaxies; we use composite model
magnitudes for A2199 galaxies.  We take this approach because of the
difference in the range of apparent magnitude covered by the faint
galaxies in the two clusters.  The SDSS web site recommends using
Petrosian magnitudes for relatively bright galaxies like those in the
SDSS spectroscopic survey.  The site recommends using composite model
magnitudes for fainter galaxies because the estimates of the Petrosian
radius are less robust at fainter magnitudes.  This difference means
that the stated absolute magnitude limits for these two clusters might
not be directly comparable.  Our primary goal here is to measure the
faint end slope of the LF.  The magnitude definition should have
only a small effect on this measurement (except for an issue with low
surface brightness galaxies discussed below).

We use NED to search for literature redshifts for all SDSS objects
that were targeted for spectroscopy but not selected by the tiling
algorithm (e.g., due to fiber collisions).  We find 20 galaxies with
redshifts that place them in Virgo.  Inspired by the ability of a
simple surface brightness cut to select probable Virgo members ($\S 2.2$), we
inspect all SDSS objects that were targeted for spectroscopy but not
selected by the tiling algorithm (e.g., due to fiber collisions).  We
find 39 LSB galaxies that are probable Virgo members and add these to
the catalog. We classify 37 of the targets satisfying the surface
brightness cut as probable members.  We classify only 2 of the higher
surface brightness galaxies as probable members despite their
significantly greater numbers.

Finally, we note that careful studies of the surface brightness
profiles of bright galaxies in SDSS show that the Petrosian magnitudes
systematically underestimate galaxy flux \citep[see discussion
in][]{dr6}.  This effect is likely caused by the pipeline
overestimating the sky background in the vicinity of large galaxies.
This underestimate decreases with increasing magnitude, but it can
still produce a $\sim$10\% underestimate of flux at $r$=16.
Unfortunately, no simple fix is effective.  Therefore, the
constraints on the LF at the bright end are probably not robust.
Because the brightest galaxies in A2199 lie at $r<14$, this issue
affects both Virgo and A2199.


\subsection{Comparison to Previous Virgo Catalogs}

Another way of determining the incompleteness of low surface
brightness galaxies in SDSS is to match the DR6 data with previous
catalogs of Virgo.  The largest of these is the Virgo Cluster Catalog
based on B band photographic plates \citep{binggeli85}, but we find
that the positional uncertainties of the VCC are too large to make
robust matches with SDSS galaxies.  Various updates to VCC positions
have been incorporated into NED.
We use the NED
positions for matching galaxies. We analyze the SDSS data for 155 VCC
objects within 1 Mpc of M87 that do not have SDSS spectroscopic
matches within 10\arcs.  Several of these objects show no obvious
counterpart in the SDSS imaging, and have no other references in NED
other than the VCC, suggesting that these are possible plate flaws in
the VCC.  Several more have likely matches with SDSS galaxies within
1$^\prime$. 
  
The more recent Virgo Photometry Catalog
\citep{young98} has better positional accuracy and magnitudes measured
in U, B$_j$, and R$_C$ bands, although it covers a much smaller area
than the VCC.  The area of the VPC is well matched to the DR6
footprint, and we find that only 26 VPC objects do not have DR6
photometric counterparts within a match radius of 10\arcs.  These 26
objects mostly result from inaccurate positions in VPC.  Three are
small galaxies in VPC which are not deblended from larger galaxies by
the DR6 pipeline (two near M84, one near M86), two are stars, and two
are galaxies not detected in DR6, one likely due to a nearby asteroid
trail.

Some recent studies have focused specifically on detecting low surface
brightness galaxies in Virgo
\citep{trentham02,trentham02b,sabatini03,sabatini05, gavazzi05}.  We explore the
properties of low surface brightness galaxies in SDSS by examining
relatively bright LSB galaxies ($B_T\leq18.5$) studied by
\citet{sabatini05}.  The magnitude cut is approximately the limit for
which (generally very blue) LSB galaxies would lie below the SDSS
spectroscopic limit.  All of the LSB galaxies are detected by the
standard SDSS pipeline, but they are subject to the ``shredding''
problem found for large galaxies.  The model magnitudes appear to be
more robust to this problem than the Petrosian magnitudes: the model
magnitudes for many galaxies are significantly brighter.  This result
motivates us to search for LSB galaxies by selecting galaxies with
$r_{Petro}-r_{cmodel}>0.3$ and $r_{cmodel}<17.47$ that would not have
been selected as spectroscopic targets.  These searches yield 53
additional probable Virgo members.

We conclude that the current SDSS photometric pipeline is insufficient
to select all low surface brightness galaxies in Virgo as
spectroscopic targets, but the pipeline usually detects these galaxies
\citep[see also][]{blanton05}.  Comparison to published photometry of
LSB galaxies in Virgo from deeper CCD imaging
\citep{trentham02,trentham02b,sabatini03,sabatini05, gavazzi05}
indicates that the SDSS pipeline magnitudes are usually within
$\sim$0.3 mag of the magnitudes from the deeper data.  The SDSS
magnitudes may not be adequate for measuring precise luminosities of
individual LSB galaxies, but they should be correct statistically for
estimating the LF.  Thus, although the SDSS spectroscopic targets
alone do not provide a complete sample of $r<17.77$ candidate Virgo
cluster members, careful treatment of the SDSS spectroscopic and
photometric catalogs enables us to recover a sample of candidate Virgo
members that is substantially more complete and probably not severely
biased against low surface brightness galaxies.

\subsection{The SDSS Virgo Cluster Catalog}

Table \ref{virgospec} lists basic photometric and spectroscopic
information for spectroscopically confirmed Virgo cluster members.
The table includes coordinates (Columns 1 and 2), $r$ magnitude
corrected for Galactic extinction (Column 3), $g-r$ color (Column 4),
$r$-band central surface brightness $\mu_{0r}$ (Column 5), redshift
(Column 6), projected distance from M87 in degrees (Column 7) and Mpc
(Column 8), and flags indicating galaxies with problematic photometry
or spectroscopy (Column 9).  Table \ref{virgonospec} lists
the photometric properties of probable Virgo members lacking reliable
redshifts.  The columns in Table \ref{virgonospec} are the same as
Table \ref{virgospec} except there is no column for redshifts.

\bibliographystyle{apj}
\bibliography{rines}


\begin{deluxetable}{lcccccccc}  
\tablecolumns{9}  
\tablewidth{0pc}  
\tablecaption{SDSS Data for Spectroscopically Confirmed Virgo Members\tablenotemark{a}\label{virgospec}}  
\small
\tablehead{  
\colhead{}    RA & DEC & $r$ & $g-r$ & $\mu_{0r}$ & $z$ & $R_P$ & $R_P$ & Flag \\
\colhead{}    (J2000) & (J2000) & mag & mag & mag &  & Degrees & Mpc &  \\
}
\startdata
 184.27314 &  12.28983 & 14.35 & 0.55 & 20.39 & 0.001340 & 3.355 & 0.992 & -- \\ 	
 184.28865 &  12.45403 & 14.21 & 0.71 & 20.51 & 0.007241 & 3.338 & 0.987 & -- \\ 	
 184.33188 &  11.94347 & 14.94 & 0.65 & 20.37 & 0.004756 & 3.328 & 0.984 & -- \\ 	
 184.36687 &  12.93234 & 16.64 & 0.22 & 21.18 & 0.006919 & 3.302 & 0.976 & -- \\ 	
 184.38049 &  11.95444 & 20.07 & 0.15 & 22.60 & 0.001316 & 3.280 & 0.970 & -- \\ 	
\enddata	     						  
\tablenotetext{a}{The complete version of this table is in the
 electronic edition of the Journal.  The printed edition contains only
 a sample.}
\tablecomments{Flags: (1) indicates problematic centroiding, often caused by blends,  (2) Galaxy light is ``shredded'' into multiple components, (3) Redshift taken from NED, (4) Alternate redshift taken from NED.
} 
\end{deluxetable}

\begin{deluxetable}{lccccccc}  
\tablecolumns{8}  
\tablewidth{0pc}  
\tablecaption{SDSS Data for Probable Virgo Members\tablenotemark{a}\label{virgonospec}}  
\small
\tablehead{  
\colhead{}    RA & DEC & $r$ & $g-r$ & $\mu_{0r}$ & $R_P$ & $R_P$ & Flag \\
\colhead{}    (J2000) & (J2000) & mag & mag & mag & Degrees & Mpc & \\
}
\startdata
 184.40237 &  12.25983 & 16.81 & 0.27 & 21.68 & 3.230 & 0.955 & 4 \\
 184.75826 &  13.98245 & 15.06 & 0.68 & 21.61 & 3.281 & 0.970 & 4 \\
 184.87540 &  13.99070 & 16.78 & 0.60 & 23.05 & 3.186 & 0.942 & 5 \\
 184.88830 &  11.01534 & 17.58 & 0.32 & 22.50 & 3.083 & 0.911 & 4 \\
 184.93511 &  12.28262 & 17.28 & 0.57 & 23.48 & 2.709 & 0.801 & 4 \\
\enddata	     						  
\tablenotetext{a}{The complete version of this table is in the
 electronic edition of the Journal.  The printed edition contains only
 a sample.}

\tablecomments{Flags: (1) indicates problematic centroiding, often caused by blends,  (2) Galaxy light is ``shredded'' into multiple components, (3) modelMag instead of petroMag, (4) No redshift available; (5) Low confidence SDSS redshift available; (6) Low confidence NED redshift available.}

\end{deluxetable}

\end{document}